# Single-photon Sagnac interferometer

G Bertocchi, O Alibart, D B Ostrowsky, S Tanzilli and P Baldi

Université Nice Sophia Antipolis, Laboratoire de Physique de la Matière Condenseée, CNRS UMR 7336, Parc Valrose 06108, Nice Cedex 2, France

E-mail: guillaume.bertocchi@unice.fr

**Abstract**

We present the first experimental demonstration of the optical Sagnac effect at the single-photon level. Using a high quality guided-wave heralded single- photon source at 1550 nm and a fibre optics setup, we obtain an interference pattern with net visibilities up to (99.2 ± 0.4%). On the basis of this high visibility and the compactness of the setup, the interest of such a system for fibre optics gyroscope is discussed.

## 1. Introduction

Experiments involving single photons are now almost usual in quantum optics laboratories, since the early works [1, 2], to the recent experiments in quantum communication [3]. However, interesting effects can still be shown involving only one photon to have a deep insight into the quantum world. The Sagnac effect [4], which produces a phase difference proportional to the angular velocity of a ring interferometer, was studied and used in optics to date only with laser [5–7], while the quantum Sagnac effect involving quantum particles has been shown only recently with superfluids or neutron beams [9, 10]. We report the first experimental demonstration of the quantum Sagnac effect using a guided-wave heralded single-photon source at 1550 nm based on spontaneous parametric down-conversion in a periodically poled lithium niobate (PPLN) waveguide and a fibre optics loop.

## 2. Theoretical description

In 1913, Sagnac demonstrated that it is possible to detect rotation with respect to inertial space with an optical system. In a ring interferometer, the rotation indeed induces a phase difference Δϕ between the two counter-propagating paths [4]. Figure 1 shows geometrically that the co-rotating wave has to propagate over a longer path than the counter-rotating one. In this figure, M and M′ stand respectively for the input and the output points of the light. At rest (case A) M and M′ are confounded; when the ring is rotating at an angular velocity Ω (case B),

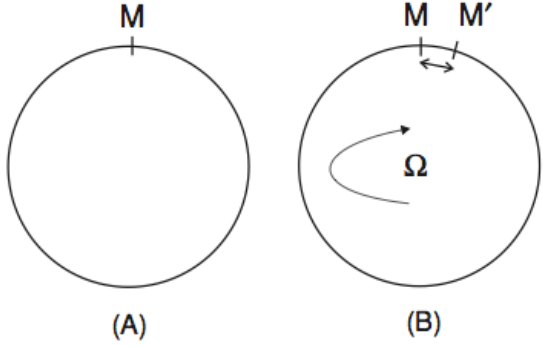

**Figure 1.** Sagnac effect in an ideal circular path. (A) system at rest; (B) system rotating at an angular velocity Ω.

M and M′ are spatially separated. The Sagnac effect, i.e. the induced phase between the two counter propagative waves, can be summed up in the following formula [8]:

$$\Delta \Phi = \frac{8\pi}{\lambda c} A\Omega, \qquad (1)$$

where $c$ is the light velocity in the vacuum, $\lambda$ the wavelength, A the interferometer area vector, and Ω is the angular velocity vector. Δϕ is thus proportional to the flux of the rotation vector Ω through the interferometer enclosed area. It is then interesting to use a multiturn optical fibre coil to enhance the Sagnac effect with multiple recirculation, each loop contributing to the total area. We can reformulate (1) in this way, considering $A$ and Ω collinear:

$$\Delta \Phi = \frac{2\pi L D}{\lambda c} \Omega, \qquad (2)$$

where $D$ is the coil diameter, $L = N\pi D$ the fibre length, and $N$ is the number of turns. We can note that Δϕ is independent of the refractive index of the fibre, as the Sagnac effect is a pure temporal delay independent of the medium [7].

We now consider the case of a single photon travelling through the interferometer. If the path length difference between the co-rotating and counter-rotating ways is shorter than the coherence length of the photon, these two paths are indistinguishable, and we therefore expect first-order interference. In practice, the input and output of the Sagnac interferometer will consist of a 50/50 beam splitter (BS), with two arms connected together by a fibre loop as depicted in figure 2. In other words, incoming photons at input 2 of the BS will emerge at one of the two outputs (labelled 3 and 4) with a probability of 1/2 . This therefore creates a quantum superposition of two possibilities, i.e. either one emerging photon takes the co-rotating path (from port 4 to port 3), or the contra-rotating path (from port 3 to port 4). The output state $|\Psi_{out}\rangle$ can thus be calculated by applying to the input state $|\Psi_{in}\rangle$ the $BS_{in}$ operator, the Sagnac operator $S$ that takes into account $\Delta\phi$, and the $BS_{out}$ operator:

$$|\Psi_{out}\rangle = (BS_{out})(S)(BS_{in})|\Psi_{in}\rangle \tag{3}$$

$$= \frac{1}{\sqrt{2}}\begin{pmatrix} 1 & 1 \\ -1 & 1 \end{pmatrix}\begin{pmatrix} 1 & 0 \\ 0 & e^{i\Delta\phi} \end{pmatrix}\frac{1}{\sqrt{2}}\begin{pmatrix} 1 & 1 \\ 1 & -1 \end{pmatrix}|\Psi_{in}\rangle. \tag{4}$$

Note that the representation used for $BS_{in}$ is one possibility among others (see [13] for more details), and since ports 3 and 4 are inverted after propagation in the loop, this operator is different when applied to an entering or outgoing photon. In our particular case, we are interested in $|\Psi_{in}\rangle = \begin{pmatrix} 0 \\ 1 \end{pmatrix}$, i.e. one photon entering the device in port 2. Then, the output state becomes

$$|\Psi_{out}\rangle = e^{-i\frac{\Delta\phi}{2}}\begin{pmatrix} \sin\left(\frac{\Delta\phi}{2}\right) \\ \cos\left(\frac{\Delta\phi}{2}\right) \end{pmatrix} \tag{5}$$

leading to a probability $\sin^2(\Delta\phi/2)$ to have the photon coming out of the interferometer in port 1 (and $\cos^2(\Delta\phi/2)$ in port 2). Repeating the experiment many times for a given $\Delta\phi$ would then lead to an expected photon number count proportional to $\sin^2(\Delta\phi/2)$ in port 1, and the complementary in the other port.

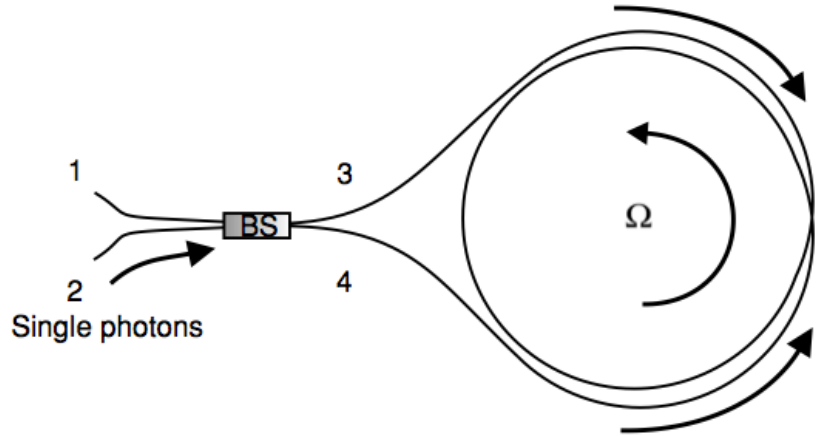

**Figure 2.** Simple scheme of the Sagnac interferometer: a beam splitter (BS) is the input and output point of the fibre loop.

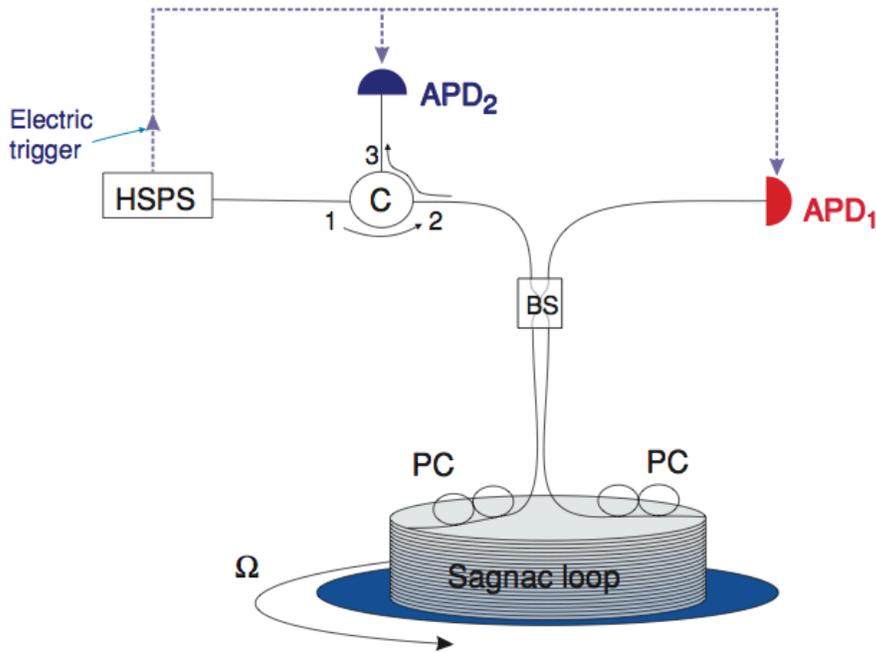

**Figure 3.** Experimental setup. Optical fibres are represented in continuous line, and electric trigger in slash-line. HSPS: heralded single-photon source. C: circulator. BS: beam splitter. APD$_{1,2}$: InGaAs avalanche photodiodes. PC: polarization controllers. The 550 meters fibre optics loop is on a plane rotating at an angular velocity $\Omega$.

## 3. Experimental setup and results

One of the main difficulties to show the Sagnac effect is that we have in theory to rotate the entire setup, which includes the source, the interferometer itself and the detection system. With a quantum optics setup including a single-photon source and avalanche photodiodes with all their electronics, this would be difficult. Since our experiment is fibred, from the output of the source to the detectors, only the Sagnac loop has to be put in rotation.

The experimental setup is detailed in figure 3. First we have a heralded single-photon source (HSPS) at 1550 nm developed in our group [11]. It is based on the creation of photon pairs by spontaneous parametric down conversion in a periodically poled lithium niobate (PPLN) waveguide [12].

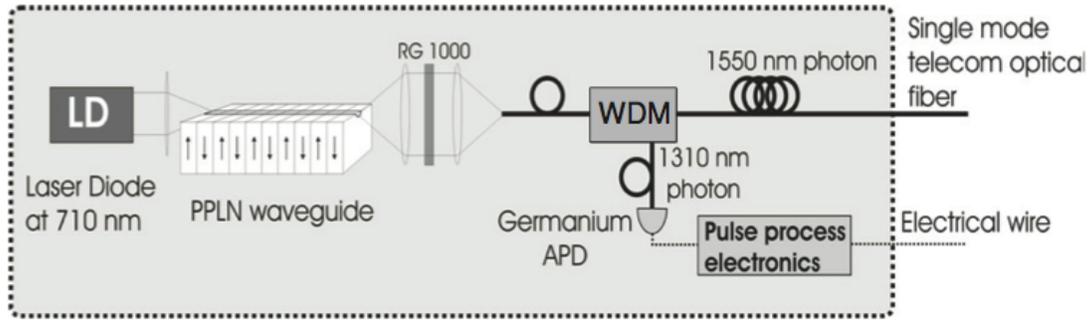

**Figure 4.** Scheme of the heralded single-photon source (HSPS). It has an electrical output which heralds the presence of a single photon in the optical fibre output. The filter (RG 1000) after the waveguide discards the remaining pump photons while the wavelength demultiplexer (WDM) separates the pairs regarding the wavelength of the photon. The 1310 nm are detected using a Ge APD in order to herald their partner photon at 1550 nm in the fibre output.

Since the two photons are emitted at the same time, the idea is to use the detection of one of them to herald the arrival of the second one. This allows us to switch on the associated avalanche photodiode (APD) during a given time-window only when the latter photon is expected. This technique enables reducing the noise of the APD dramatically. In our configuration, for a PPLN period of 13.6 µm, a pump photon at 710 nm (from a CW laser) is converted into a pair of photons, whose wavelengths are centred at 1310 nm and 1550 nm. The 1310 nm photon is detected with a germanium APD working in continuous mode (quantum detection efficiency $\eta_{Ge} = 0.06$). This detection heralds the arrival of the complementary 1550 nm photon which is detected using two InGaAs APDs ($\eta_{InGaAs} = 0.1$) working in gated mode with 5 ns detection time windows (see figure 4 for additional details).

Note that our InGaAs APDs show a typical probability of darkcount of $5 \times 10^{-5}$ ns$^{-1}$. For a pump power of 10 µW at 710 nm, the HSPS delivers into a connectorized standard single mode optical fibre heralded single photons with the following probabilities: P (0) = 0.81, P (1) = 0.17, P (2) = $5 \times 10^{-3}$, where $P(0)$, $P(1)$ and $P(2)$ stand for the probability to have a time-window containing 0, 1 and 2 photons respectively (the probability of having more than 2 photons being negligible). More details about this HSPS are given in [12]. Typically, the average rate of heralds is $10^5$ s$^{-1}$, limited by the Ge APD (which saturates above 120 kcounts s$^{-1}$). This rate corresponds to an average rate of 1550 nm heralded photons injected into the loop of $2 \times 10^4$ s$^{-1}$. This means that each heralded photon is in average alone in the interferometer during its propagation time (2.67 µs).

Once emitted, the single photons are sent to the fibre loop via a circulator (C), and a BS as described in theoretical part, which stands for the input and output of the Sagnac loop. Photons that go out in port 1 are directly counted in APD$_1$. The ones that go out in port 2 are sent to APD$_2$ thanks to the circulator. Both APDs have the same characteristics and are electrically triggered by the HSPS only when photons at 1550 nm are expected, as explained before. The fibre optics loop is on a rotating plane whose angular velocity $\Omega$ is controlled. As a consequence of this configuration, the rotation applied on the loop induces motion and twist of the fibre, hence constraints, that cause a change on the guided mode polarization. Thanks to the polarization controllers (PC) in the loop, we find a polarization configuration for which the loop acts as a mirror when it is motionless, and which is not affected by a rotation of some turns. Actually, we measure with a standard laser that the output power of each arm is stable within $10^{-3}$, i.e. the polarization state is not affected during the entire rotation.

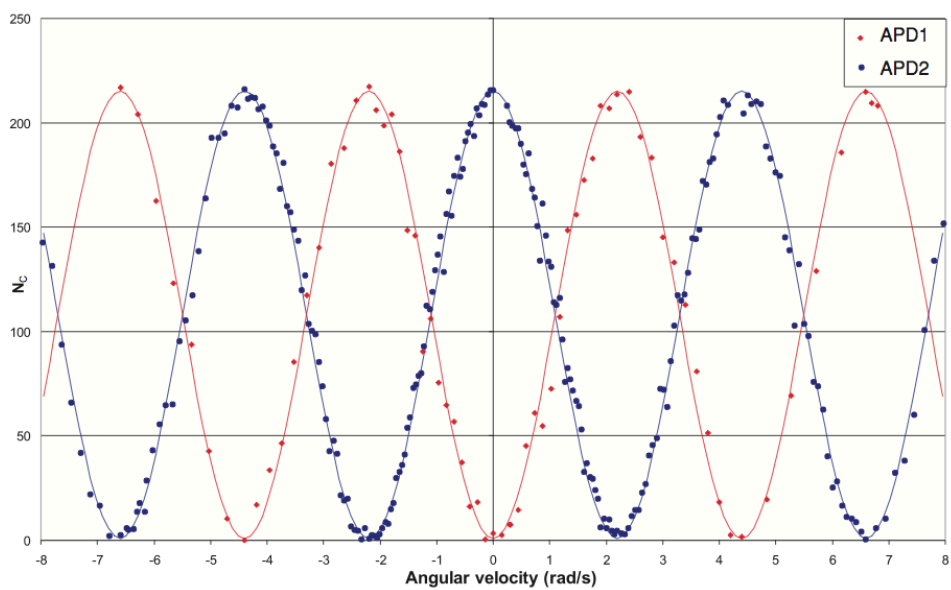

**Figure 5.** Experimental results. The interference pattern shows the Sagnac effect with visibility up to (99.2 ± 0.4)%. Each point corresponds to the photon number count $N_c$ for a 300 ms integration time at a given angular velocity $\Omega$. We can note that $\Omega_{n\pi} = n2.2$ rad s$^{-1}$, corresponding to formula (2).

An important parameter is $\Omega_\pi = \lambda_c/2LD$, corresponding to $\Delta\phi = \pi$. In our setup, we use a 550 m long fibre, rolled up around a 20 cm spool. With this configuration, $\Omega_\pi = 2.2$ rad s$^{-1}$ at $\lambda = 1550$ nm. The rotating plane allows the motion of the Sagnac loop, with an angular velocity up to $\Omega_{max} = 10$ rad s$^{-1}$. To access different angular velocities, a good approximation is to start from $\Omega_{max}$ and let the velocity vanish, due to the friction force. The angular velocity is measured using an infrared light barrier and the total recording time is around 1 min, corresponding to 40 turns. In order to detect the interference, we count the number of photons $N_c$ during a given integration time. Each point of the plot in figure 5 corresponds to an integration time of 300 ms, and hence to a precision on the angular velocity $\Omega$ of 0.1 rad s$^{-1}$. For each point we have subtracted the constant noise due to the dark counts (25 s$^{-1}$ in average). Moreover, in order to lower the fluctuations, an averaging over five records has been made. This leads to interference patterns with net visibilities up to 99.2 ± 0.4%, along with evident complementarity between the two outputs of the Sagnac interferometer.

## 4. Discussion

The classical Sagnac interferometer using fibre optics loops is at the heart of the fibre optics gyroscopes (FOG). An important drawback in FOG is due to the nonlinear optical Kerr effect. Using a laser, an imbalance in the power level of the two contra-propagating waves can cause a nonreciprocal phase difference [8, 10]. This can be avoided by reducing the power in the fibre. In this way, using a single-photon source represents the minimal power we can use and the single-photon Sagnac interferometer cancels the nonlinear effects.

On the other hand, in a classical FOG a too low power leads to excess shot noise. It is then interesting to analyse the effect on the shot noise limitation in our experiment. To simplify, let us consider that we have perfect detectors ($\eta = 1$) and that the electromagnetic field can be described as a one-photon Fock state ($P(1) = 1$), eliminating photon shot noise. However, in order to carry out our phase measurement we have to repeatedly count the single photons incident on one or the other of the two detectors. This leads to binomial statistics with a standard deviation $\sigma_N$ proportional to the square root of the number of photons $N$. This is the same result as that obtained for the shot noise with a classical measurement but has another underlying cause.

Finally, we must emphasize that with the present technology, we think it is not interesting to consider a FOG based on single photons. For instance, the sensitivity (i.e. the most little phase difference $\Delta\phi$ that can be detected) of a classical gyroscope is of the order of 1 μrad [8]. The most little phase difference that a single-photon gyroscope could detect is limited by the standard deviation $\sigma_{\Delta\phi}$, which is given by

$$\sigma_{\Delta\phi} = \frac{\sigma_N}{N} = \frac{1}{\sqrt{2N}} \qquad (6)$$

where the factor 2 comes from the fact that the gyroscope working point is $\Delta\phi = \pi/2$, i.e. at half the maximum number of photon $N$. In the best case, current APDs would limit $N$ to $10^7$ s$^{-1}$. A microradian resolution would then require an integration time of around $5 \times 10^4$ s (14 h!), a too long time for a realistic FOG.

## 5. Conclusion

We have presented the first experimental demonstration of the Sagnac effect using light at the quantum level. The homemade heralded single-photon source at 1550 nm we use is based on guided-wave spontaneous parametric down conversion in a periodically-poled lithium niobate waveguide. The use of a fibred interferometer and the control of the polarization state allow us to let the source, detectors and electronics motionless. This double path quantum interference shows a fringe visibility greater than 99%. While this setup cancels the detrimental nonlinear effects present in fibre optics gyroscope, the required integration time seems to limit the interest of the single-photon Sagnac interferometer for such gyroscopes.